\documentclass[preprint2]{aastex}

\slugcomment{Not to appear in Nonlearned J., 45.}

\shorttitle{ENB from environment of GRB jets}
\shortauthors{W. Bednarek \& A. \'Smia\l kowski}

\begin{document}


\title{Contribution to the extragalactic neutrino background from  dense environment of GRB jets}


\author{W. Bednarek \& A. \'Smia\l kowski}
\affil{University of Lodz, Faculty of Physics and Applied Informatics, Department of Astrophysics, 90-236 Lodz, ul. Pomorska 149/153, Poland}

\email{wlodzimierz.bednarek@uni.lodz.pl; andrzej.smialkowski@uni.lodz.pl}



\begin{abstract}
Long gamma-ray bursts (GRBs) are at present well confirmed sites of acceleration of particles to relativistic 
energies due to observations of gamma-ray emission in the GeV-TeV energy range.  
We consider a scenario in which the mechanism accelerating electrons is also responsible for acceleration 
of hadrons in the GRB jets to multi-PeV energies.  
Since progenitors of long GRBs are massive stars still immersed in dense stellar clusters,
these hadrons can efficiently interact with the matter after escaping from the jet. 
We calculate the spectra of neutrinos from the interaction of those hadrons with the matter of a huge cloud 
surrounding GRB. Those neutrinos form an afterglow on a time scale determined by their diffusion time scale 
in the cloud. Due to this long delay, their identification with any specific GRB is not possible. 
We estimate contribution of neutrinos from such afterglows to the extragalactic neutrino background 
recently reported by the IceCube Observatory.
\end{abstract}




%
%
\section{Introduction}

Long gamma-ray bursts are evidently sites of violent non-thermal processes occuring within their jets and/or 
surrounding environment (see recent review by Zhang 2019). In fact, evidences of the existance of the hard 
gamma-ray emission from GRBs, in the GeV energy range obtained with the EGRET and the {\it Fermi}-LAT telescopes 
(e.g. Hurley ey al. 1994, Ackermann et al. 2013, Ackermann et al. 2014),
have been recently confirmed with the observations of the TeV $\gamma$-ray emission by Cherenkov telescopes 
(Acciari et al. 2019a, Abdalla et al.2019). This GeV-TeV gamma-ray emission is believed 
to be produced by relativistic electrons accelerated in a relativistic jet (Acciari et al. 2019b), but 
acceleration of hadrons in a similar mechanism seems to be very likely.
If this hard GeV gamma-ray emission comes from the region co-spacial with observed keV emission, then 
the condition, on not beeing internally absorbed, indicate the minimum Lorentz factor of the jet in the case of 
GRB 080916C of the order of $\sim$870 (Abdo et al~2009). It is at present commonly expected that long GRBs 
are due to the asymmetric explosions of massive stars (so called "collapsar" model proposed by 
Woosley et al. 1993 and Paczy\'nski 1998). Their progenitor stars have to be still surrounded by 
a dense matter of giant clouds due to very short evolutionary periods of supermassive stars.  
 
Observed non-thermal radiation from GRBs clearly strongly argue for the presence of relativistic hadrons. 
They can be the
main contributors of neutrinos at a broad range to the Extragalactic Neutrino Background (ENB) which has been 
recently discovered by the IceCube neutrino telescope (Aartsen et al. 2015, Abbasi et al. 2021). However,
 correlation studies of the IceCube neutrino events and presently observed GRBs do not report coincidences 
between prompt $\gamma$-ray emission and neutrino events (Aartsen et al. 2017). This puts into question 
the importance of hadronic processes in GRBs. In the present work, we propose a scenario in which direct
link between GRBs and neutrino events is in fact not expected, in spite of the efficient acceleration of hadrons 
in jets of GRBs. We argue that hadrons, accelerated in relativistic jets of GRBs can efficiently escape from 
the jet into surrounding dense cloud. Due to a relatively low 
interaction rate of hadrons with the background matter (and their large diffusion distance, of the order of 
a few tens of parsecs), hadrons should
produce delayed neutrino afterglows which cannot be detected instantly after occurence of a specific GRB. 
However, neutrinos, from the whole population of GRBs in the Universe, can 
contribute significantly to the observed ENB. In the considerred model, neutrinos are produced on a time 
scale of a tausend of years after appearace of the GRB, in contrust to the "classical" afterglow scenario 
(e.g. Waxman \& Bahcall 2000) 
for the non-thermal emission from GRBs which usually lasts on a months time scale.  

\section{GRB within dense open cluster}

Long duration GRBs are expected to be final products of asymmetric explosions of short lived very massive 
stars ($>$30 M$_\odot$). These progenitors of GRBs are not able to move in such a short period (of the order 
of $\sim$Myrs) outside the parent dense cluster. Therefore, they are expected to be still immersed in dense gas 
before explosions as supernovae. In fact, a large population of such young and compact clusters in the nearby 
galaxies (with ages below 100 Myrs, masses above $10^6$ M$_\odot$ and radii $<$15 pc) has been found by using 
the Hubble Space Telescope (HST) (see e.g. Holtzman et al. 1992, Whitemore \& Schweizer 1995, Miller et al. 
1997, Zeft et al. 1999). At farther distances, the proto-globular clusters are also observed, with masses 
in the range $(1-20)\times 10^6$ M$_\odot$ and sizes of a few tens of pc (Vanzella et al.~2016). It has been 
suggested that such young clusters can continuously form in the local Universe as a result of galaxy mergers 
(Schweizer 1987). Massive stars can also produce fast and dense stellar winds which are confined by 
the surrounding cloud. 
They can also lose matter explosively forming more or less uniform cocoon. In fact, in the case of the most
 massive binary system observed in the Galaxy, Eta Carinae (companion stars with masses 
$100$~M$_\odot$ and $30$~M$_\odot$), a dense cloud  with the mass of 
$\sim$12 M$_\odot$ and radius $3.4\times 10^{17}$~cm is observed (Smith et al. 1998, Smith et al.~2003). 
For the above mentioned parameters, we estimate that Eta Carinae is submerged within the region with average 
density of the order of $\sim$2$\times 10^4$~cm$^{-3}$.

GRB jets have to propagate through the wind bubbles formed as a result of the interaction of the progenitors 
winds with the surrounding medium.     
We investigate the scenario in which the GRB jets propagate in dense medium of a huge cloud surrounding
projenitor star (see Fig.~1). Hadrons are expected to be accelerated within the relativistic jets of GRBs. 
Moreover, energies of hadrons, escaping to the medium from the jet, are relativistically busted with 
the Lorentz factor of the jet. 
After leaving the wind cavity, hadrons interact with the dense medium of a huge cloud producing high energy
neutrinos through decay of pions. Hadrons are either confined within the cloud, 
losing efficiently energy, or they escape from the cloud with significant energy losses. 
The lower energy hadrons produce neutrinos. On the other hand, the higher energy hadrons diffuse throught
the cloud without significant energy losses, contributing 
to the extragalactic cosmic ray background. Since the interaction time scale of protons takes typically 
thousands of years, observed neutrino events cannot be
identified with any recently detected GRB. Those neutrinos form a neutrino afterglow which is delayed
on a time scale of thousend of years. The aim of this article is to estimate the contribution of neutrinos, 
produced in terms of such scenario, to the extragalactic neutrino background in the Universe.

\begin{figure}
\centering
\includegraphics[width=.5\textwidth]{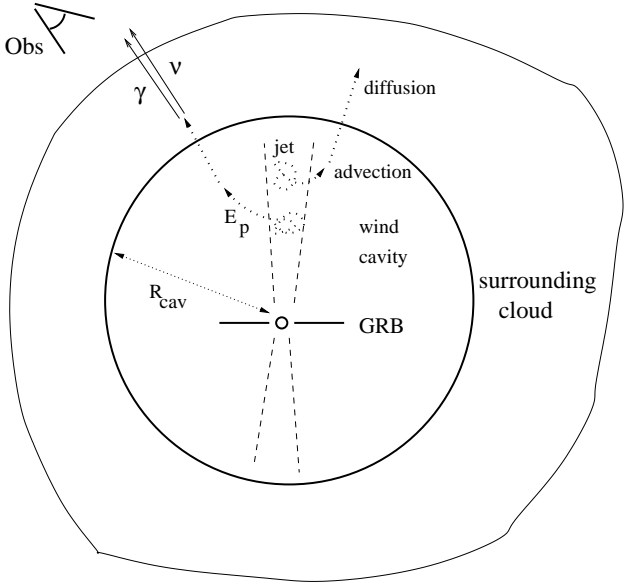}
\caption{Schematic presentation of the long GRB which is produced in explosion of a massive Wolf-Rayet type star
within huge and dense cloud. The progenitor star produces a stellar wind cavity with the radius, $R_{\rm cav}$, 
within which the relativistic jet propagates. Protons, accelerated in the outer parts of the jet, escape from 
it into the stellar wind region. In the wind cavity, protons either lose energy on the adiabatic process 
(due to the expansion of the stellar wind) or move balistically through the cavity. They are injected into 
the dense giant cloud. During diffusion process in the cloud, relativistic  protons collide with the matter 
producing high energy neutrinos.}
\label{fig1}
\end{figure}

\section{Acceleration of hadrons in the jet}

Relativistic jets of GRBs emit clearly non-thermal radiation. Therefore, they are expected to provide good 
conditions for the acceleration of particles. 
We assume two cases regarding the homogenity of the medium  surrounding the GRB. In the first case, 
the medium uniform. In the second case, it is non-uniform with density scaled with power law  function of 
the distance from the central engine $n \propto R^{-k} $ 
with $k=2$ (Dai \& Lu~1998).  We consider the decelerating jet of the GRB which
Lorentz factor (in the observer's reference frame) evolves in time as,
\begin{eqnarray}
\Gamma (t) = \Gamma_{\rm 0} (t/t_{\rm 0})^{-(3-k)/(8-2k)},
\label{eq1}
\end{eqnarray}
where $\Gamma_{\rm 0} = 500\Gamma_{\rm 2.7}$ is the initial jet Lorentz factor at the time $t_{\rm 0}$ (in 
seconds). This Lorentz factor of the jet is expected to be of the order of a few hundred for the total jet 
energy of the order of  $\sim 10^{55}$ erg and the density of surrounding medium in the range 
$n = 0.1-10^4$ cm$^{-3}$ (Sari et al.~1997, see Eq.~14). 
In the case of decelerating jets, the relation between the distance from the jet base and the time, in 
the observer's reference frame, can be obtained by integrating the relation $dR = 2c[\Gamma(t)]^2dt$. Then, 
\begin{eqnarray}
R = 8c\Gamma_{\rm 0}^2t_{\rm 0}(t/t_{\rm 0})^{1/(4-k)} = R_{\rm 0}(t/t_{\rm 0})^{1/(4-k)}, 
\label{eq2}
\end{eqnarray}
and $R_{\rm 0} = 8c\Gamma_{\rm 0}^2t_{\rm 0}\approx 6\times 10^{16}t_{\rm 0}\Gamma_{2.7}^2$~cm.  
The Lorentz factor of the jet evolves with the distance from the jet base according to,
\begin{eqnarray}
\Gamma (R) = \Gamma_{\rm 0} (R/R_{\rm 0})^{-(3-k)/2}. 
\label{eq3}
\end{eqnarray}
The maximum energies of protons, in the
jet reference frame at the distance, $R$, from its base, are defined by their energy gains from the 
acceleration mechanism. Since the acceleration process in relativistic jets is not at present well known 
phenomena, the acceleration time scale is often parametrized by a simple formula related to 
the Larmor radius of the particle, 
\begin{eqnarray}
\dot{E}_{\rm acc} = cE/(\eta R_{\rm L}) \approx 10^{12}B/\eta_1~~~{\rm eV/s}, 
\label{eq4}
\end{eqnarray}
where $\eta = 10\eta_1$ is the so called acceleration parameter, and 
$R_{\rm L} = E/eB\approx 3\times 10^{-3}E/B$~cm is the Larmor radius and $E$ is the proton energy in eV 
and $B$ is the magnetic field strength in the jet (in Gauss)  at the distance, $R$, from its base. 
The magnetic field can be estimated assuming that it is generated locally in the jet. Following 
Razzaque et al.~(2013), we relate the magnetic field to the local parameters of the jet by,
\begin{eqnarray}
B(R) = \left({{2L_{\gamma, iso}}\over{c}}\right)^{1/2}\times \frac{\beta}{R\cdot\Gamma(R)}
~~~{\rm Gs},
\label{eq5}
\end{eqnarray}
where $\beta = (\varepsilon_{\rm B}/\varepsilon_{\rm e})^{1/2}\approx 0.14$, $\varepsilon_{\rm B}\sim 0.001$ 
is a fraction of the shock energy that is carried by the magnetic field, $\varepsilon_{\rm e}\sim 0.1$ is 
a fraction of the shock energy that is carried by the relativistic electrons (see Piran~2005, Panaitescu \& Kumar~2001, Yost~et~al.~2003), 
$L_{\gamma, iso}\approx L_{\rm 0}(t_{\rm L}/t(R))^{\delta}$ is the isotropic-equivalent 
$\gamma$-ray luminosity, $L_{\rm 0} = 10^{52}L_{52}$ erg/s is the peak luminosity at the time $t_{\rm L} = 10$~s
 and 
the index $\delta = 1.17$ is applied as observed in the GRB 130427A. 
For the applied scaling parameters and the Lorentz factor of the jet at one second after initial flash 
equal to 500, the magnetic field strength in the jet of GRB 130427A is estimated to be of the order of a gauss 
at the distance at which protons escape from the jet.

On the other hand, we assume that protons lose mainly energy on the pion production in collisions with 
radiation and on the adiabatic expansion of the emission region within the jet.
The adiabatic energy loss rate is defined as 
\begin{eqnarray}
\dot{E}_{\rm ad} = c E\Gamma/R.
\label{eq6}
\end{eqnarray}
By comparing the energy gains with the energy losses on the adiabatic process, we obtain the maximum energies 
of accelerated protons,
\begin{eqnarray}
E_{\rm ad}(R)\approx 4\times 10^{3}(\Gamma_{2.7}t_{\rm 0}B(R)/\eta_1) 
(R/R_{\rm 0})^{\frac{5-k}{2}}~~~{\rm TeV}.
\label{eq7}
\end{eqnarray}

However, close to the base of the jet where the density of radiation produced in the jet is high, 
accelerated protons can also efficiently lose energy in collisions with photons on the pion production. 
We estimate the distance, from the base of the jet, at which this energy loss process becomes negligible. 
The optical depth for protons on the pion production in collisions with non-thermal radiation from the jet is
estimated from, 
$\tau_{\rm p\varepsilon} =  \sigma_{\rm p\varepsilon}n_{\varepsilon}R/\Gamma$, 
where the cross section for this process is $\sigma_{\rm p\varepsilon} = 5\times 10^{-28}$~cm$^2$, and $n_{\varepsilon}$ is the number density 
of photons in the co-moving frame of a GRB jet estimated from (see e.g. Razzaque, S., 2013),
$n_{\varepsilon} = d_{\rm L}^2F_{\varepsilon}(t)/[R^2c\Gamma \varepsilon_{\rm min}(1+z)]$, 
where $F_{\varepsilon}(t)$ is the observed hard X-ray photon flux at the time "t" after the initial flash. 
This flux is assumed to vary as observed in the case of GRB 130427A according to $F_{\varepsilon}(t) = F_{\rm 0}(t/t_0)^{-\delta}$, where $\delta = 1.17$ (Ackermann~et~al.~2014). The characteristic photon energy in the observer's reference frame, $\varepsilon_{\rm min}$, 
for which pion production process can occur is estimated from $\varepsilon_{\rm min} = 
\varepsilon_{\rm th} \Gamma/\gamma_{\rm p}(1+z)$, where $\varepsilon_{\rm th}\approx 140$ MeV, and 
the Lorentz factor of relativistic proton is $\gamma_{\rm p} = E_{\rm ad}/m_{\rm p}c^2$. 
We estimate the distance, $R_{\rm loss}$, from the base of the jet at which
the energy losses limit the acceleration process of protons to energies below $E_{\rm ad}$. 
This  distance scale corresponds to 
the optical depth for pion production, $\tau_{\rm p\varepsilon}$, equal to unity.

The spectrum of accelerated protons is assumed to be well described by a power law function, 
$dN_{\rm p}/dE = A_{\rm p}E^{-2}$, where $A_{\rm p}$ is the normalization coefficient. 
If the interaction of relativistic protons in the jet is effective, then the injection spectrum is modified
by the process of proton energy losses already in the jet. Therefore, the injected spectrum of protons takes 
the form, 
\begin{eqnarray}
dN_{\rm p}/dE = A_{\rm p}E^{-2}exp(-\tau_{\rm p\varepsilon}(E)).
\label{eq9}
\end{eqnarray}
The coefficient $A_{\rm p}$  is obtained from the normalization of the proton spectrum to a part,
$\varepsilon_{\rm p}$, of jet power which is assumed to be equal to 
$L_{\rm \gamma, jet} = L_{\rm \gamma,iso}\alpha^2/2\varepsilon_{\rm e}$.
$\varepsilon_{\rm p}$ is assumed to be equal to $3\%$ of the fraction of the power emitted from the jet in $\gamma$-rays, and $\varepsilon_{\rm e} = 0.1$ is the power in relativistic electrons. 

Protons, accelerated at a specific distance, $R$, from the base of the jet, start to escape effectively from the jet when their diffusion distance (during the adiabatic time scale, $\tau_{\rm ad} = E/\dot{E}_{\rm ad}$) becomes comparable to the perpendicular extend of the jet, i.e. $Z_{\rm dif} = 0.1\alpha_{-1}R$, where 
$\alpha = 0.1\alpha_{-1}$ rad is the jet opening angle. In the case of the Bohm diffusion, this distance is given by $Z_{\rm dif} = \sqrt{2D_{\rm dif}\tau_{\rm ad}}$, and the Bohm diffusion coefficient is $D_{\rm dif} = cR_{\rm L}/3$. Then, protons escape with energies,
\begin{eqnarray}
E_{\rm esc}(R)\approx 1.5\times 10^{8}\alpha_{-1}^2\Gamma_{2.7}^3t_{\rm 0}B(R)\cdot (R/R_{\rm 0})^{-(1-k)/2}~~~{\rm TeV}.
\label{eq10}
\end{eqnarray}
By comparing the maximum allowed energies of protons, $E_{\rm ad}$, at a specific distance from the jet base, $R$, with their escape energy, $E_{\rm esc}$, we obtain the distance above which locally 
accelerated protons start to effectively leakage from the jet into the surrounding medium,
\begin{eqnarray}
R_{\rm esc} = R_{\rm 0}(3\times 10^5\Gamma_{2.7}^2\eta_1\alpha^2_{-1}/8)^{1/(3-k)}.
\label{eq11}
\end{eqnarray} 
We assume that the process of proton acceleration becomes inefficient when the jet becomes sub-relativistic, i.e. $\Gamma (R) = 1.1$. This happens at the distance 
$R_{\rm max} = R_{\rm 0}(\Gamma_{\rm 0}/1.1)^{2/(3-k)}$.
Relativistic protons accelerated at a specific distance, $R$, are advected with the jet plasma flow. They  loose energy on the adiabatic process up to the moment of their escape from the jet.
Note that, the energies of protons, $E_{\rm jet}$, escaping from the jet to the progenitor star wind region, are additionally boosted by the Lorentz factor of the jet. 

\begin{figure}
\centering
\includegraphics[width=.55\textwidth]{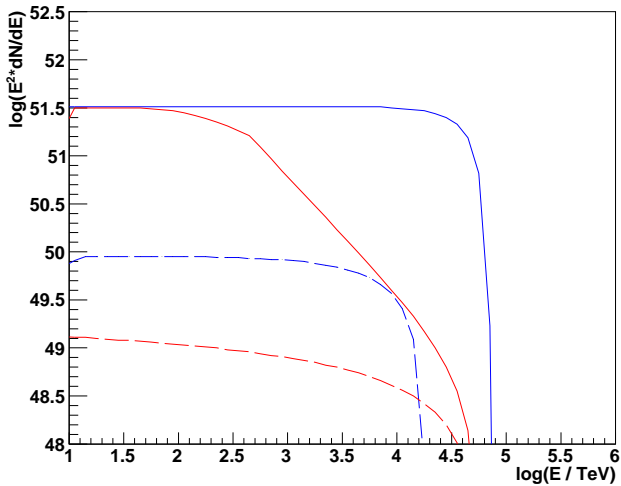}
\caption{Spectral energy distribution (SED) of relativistic protons, which escaped from the progenitor star wind region into the giant cloud, is shown for the two models of the surrounding medium homogenity, i.e. the uniform medium (thick curves) and non-uniform medium (thin curves). The model with a negligable adiabatic losses is shown by the solid curves and with adiabatic energy losses (as defined in Sect. 4.1) is shown by 
the dashed curves.}
\label{fig2}
\end{figure}

\section{Propagation of hadrons around GRB}

Protons escape from the jet at first into the massive star wind cavity
and latter  into the giant cloud in which the massive projenitor star exploded.
Below we discuss the conditions for the propagation of protons in these two regions.
We argue that a significant amount of protons, injected from the jet, lose only a part of their energy due to the adiabatic process in the expanding GRB progenitor wind. The interaction of protons with the matter of the wind is inefficient since the density of the wind at the parsec distance scale is rather very low, 
\begin{eqnarray}
n_{\rm w}\approx 3.7\times 10^{-3}\dot{M}_{-6}/(R_{\rm pc}^{2}v_3)~~~{\rm cm}^{-3},
\label{eq12}
\end{eqnarray}
where the wind velocity $v_{\rm w} = 1000v_3$ km~s$^{-1}$, and the mass loss rate $\dot{M}_{\rm WR} = 10^{-6}M_{-6}$ M$_\odot$~yr$^{-1}$, and  
$R = 1R_{\rm pc}$ pc is the distance from the star in parsecs. On the other site,  
protons can lose significant amount of their energy in collisions with the background matter in the giant molecular cloud surrounding the progenitor star. Therefore, we concentrate on the calculation of the neutrino spectra produced by these relativistic protons with the matter of the huge and dense cloud.

\subsection{Propagation of hadrons in the wind cavity}

The conditions within the stellar wind cavity change significantly with the distance from the massive star, 
a progenitor of the GRB. Let us consider the typical parameters of the WR type star mentioned above. Its surface magnetic field strength is fixed on $B_{\rm WR} = 10^3B_3$ Gs.
At large distance from the star, the magnetic field is estimated as,
$B(R) = 3\times 10^{-5}B_3/R_{\rm pc}~~~{\rm Gs}$.
Depending on the energy of protons, they can be either captured by the magnetic field or escape almost ballistically through the wind. In the first case, protons are expected to suffer adiabatic energy losses in the expending wind. In the second case,
they move freely to the giant cloud region without significant adiabatic energy losses. We estimate the energies of relativistic protons below which they are frozen in the stellar wind by comparing their Larmor radius with the distance from the star. It is found that protons with energies, 
$E_{\rm bal} < 3\times 10^4~{\rm TeV}$, are frozen into the GRB progenitor wind.
But, the largest energy protons escape from the wind region balistically, without significant energy losses  due to the adiabatic expansion of the stellar wind.
The energies of protons, after losing energy on the adiabatic process in the wind cavity, can be determined from $E_{\rm cav} = E_{\rm fin}R_{\rm fin}/R_{\rm cav}$,
where $R_{\rm cav}$ is the radius of the cavity filled with the stellar wind. $R_{\rm cav}$ depends on the age of the star,
its wind parameters and the parameters of the surrounding giant cloud (see Weaver~et~al.~1977), according to
$R_{\rm cav} = 18 (M_{-6}V_3/n_2)^{1/5}t_3^{3/5}~~~{\rm pc}$,
where $t = 3t_3$ Myr is the age of the star, and $n_{\rm cl} = 100n_2$~cm$^{-3}$ is the density of the giant cloud.

In Fig.~2, we show the proton spectra, escaping from the stellar wind region into the giant cloud, for the 
three different models for the energy losses of protons: (A) adiabatic energy losses, both within the jet and stellar wind, are taken into account; (B) adiabatic losses important only in the jet; (C) adiabatic losses not
important. Those three models are considered since it is not to the end clear whether adiabatic losses of protons play any role during their propagation. As an example, we use the following parameters for the considered scenario: $\Gamma_{\rm 0} = 500$, $L_0 = 10^{52}$~erg~s$^{-1}$, $\varepsilon_{\rm B}= 10^{-3}$, 
$\alpha = 0.1$, $\eta_1=10$. It is evident that adiabatic losses of protons, if important can significantly extract energy from relativistic hadrons. Therefore, we expect that interesting fluxes of neutrinos should be produced only in the case of a free escape of protons into the surrounding dense cloud.

\subsection{Propagation of hadrons in the cloud}

Let us consider a giant molecular cloud with the example parameters: the radius, 
$R = 30R_{\rm 30}$~pc and the mass $10^6M_6$~M$_\odot$ in which the GRB has been exploded. Then, protons, escaping from the jet, have to propagate through this cloud. The average density of such cloud can be estimated on  
\begin{eqnarray}
n_{\rm cl}\approx 270M_6/R_{30}^3~cm^{-3}. 
\label{eq14}
\end{eqnarray}
We assume the magnetic field in the cloud is of the order of $B_{\rm cl} = 10^{-4} B_{-4}$~G. The Bohm diffusion coefficient of relativistic protons is then
\begin{eqnarray}
D_{\rm dif} = R_{\rm L}c/3\approx 3\times 10^{23}E_{\rm TeV}/B_{-4}~~~{\rm cm^2~s^{-1}}. 
\label{eq15}
\end{eqnarray}
The diffusion time scale of protons through the cloud is,
\begin{eqnarray}
T_{\rm dif} = R_{\rm cl}^2/2D_{\rm dif}\approx 1.35\times 10^{17}R_{\rm 30}^2B_{-4}/E_{\rm TeV}~~~{\rm s}.
\label{eq16}
\end{eqnarray}
Then, the interaction rate of protons in the cloud is given by
$T_{\rm int} = (c n_{\rm cl}\sigma_{\rm pp})^{-1} \approx 4\times 10^{12} R_{30}^{3}/M_6~{\rm s}$.
We follow numerically the propagation of injected protons in the cloud taking into account their energy losses on the  pion production process in collisions with the matter of the cloud with the aim to calculate the spectra of VHE neutrinos.

\section{Delayed VHE neutrinos}

We have simulated spectra of mesons produced by relativistic protons at a given energy using the CORSIKA Monte Carlo package (Heck et al.~1998). Then, the spectra of neutrinos, produced by relativistic protons during their diffusion and collisions with the matter of the cloud, are obtained. The effects of energy losses of protons during their propagation within the jet and the wind region of progenitor star are taken into account as discribed above. 
Protons are assumed to be injected locally in the jet with the power law spectrum and normalization described above. The multiple interactions of the lower energy protons with the matter of the cloud are also taken into account. In such a way, we are able to calculate the neutrino spectra from the GRBs immersed in dense clouds with arbitrary parameters. As a last step, we estimate the contribution of the whole population of GRBs in the Universe to the neutrino extragalactic background.

\subsection{Extragalactic Neutrino Background from GRBs}

We calculate diffuse neutrino background from the whole population of GRBs taking into account the redshift rate of the GRBs, i.e. $R_{GRB}(z)$ (see e.g. Murase~2017).
We compute the neutrino fluxes as a function of the redshift, $z$, taking into account the redshift dependence of the jet opening angle and the isotropic equivalent gamma-ray luminosity of GRBs (basedon the redshift dependences presented in Lloyd-Ronning~et~al.~2019).

The differential flux of extragalactic neutrinos from GRBs is calculated from the formula,
\begin{eqnarray}
{{dN}\over{dE_\nu dt dS d\Omega}} = \phi = 
\frac{c}{4\pi H_{0}}\times \\ \nonumber
\times \int_{0}^{z_{max}}dz \frac{R_{GRB}(z)}{F(z,\Omega_{m}, \Omega_{\Lambda})}\cdot\frac{dN_{\nu}(E'_{\nu})}{dE'_{\nu}}(1+z)^{-3/2},
\label{eq00}
\end{eqnarray} 
where $E'_{\nu} =(1+z)E_{\nu}$ is the energy of neutrinos at redshift $z$ and $E_\nu$ is the energy at the observer, $H_0=71\, {\rm km} \,{\rm s}^{-1}\, Mpc^{-1}$, $z_{max}=10$
and $F(z,\Omega_{m}, \Omega_{\Lambda}) =\sqrt{\Omega_{\Lambda}(1+z)^{-3} +\Omega_{m}}$, with adopted 
values for $\Omega_{\Lambda}=0.7$ and $\Omega_{m}=0.3$.

On Fig~3, we show the extragalactic diffuse neutrino background produced in GRBs for the case of three considered models. Those models consider ina different way the importance of the adiabatic energy losses of hadrons propagating around the GRB progenitor. We conclude that the significant contribution to the extragalactic neutrino background is obtained in the case of a negligible adiabatic energy losses of relativistic hadrons in the projenitor star wind region. 
In such a case, our scenario is able to contribute significantly to the ENB at energies below 
$\sim$100 TeV. However, the ENB at higher energies should be produced in another model, either in the inner part of the relativistic jets of GRBs or in other type of cosmic sources (e.g. active galactic nuclei, starburst galaxies, supernova remnants, ...).  

Our model bases on the assumption that the efficiency of acceleration of hadrons is equal to the efficiency of acceleration of leptons which is responsible for the observed 
$\gamma$-ray power of the GRB. Therefore, prediced neutrino fluxes should be even enhanced in the case of a more efficient acceleration of hadrons in comparison to leptons in jets of GRBs. Such difference might be related to the injection problem of particles (hadrons/leptons) into the acceleration mechanism.

\begin{figure}
\centering
\includegraphics[width=.55\textwidth]{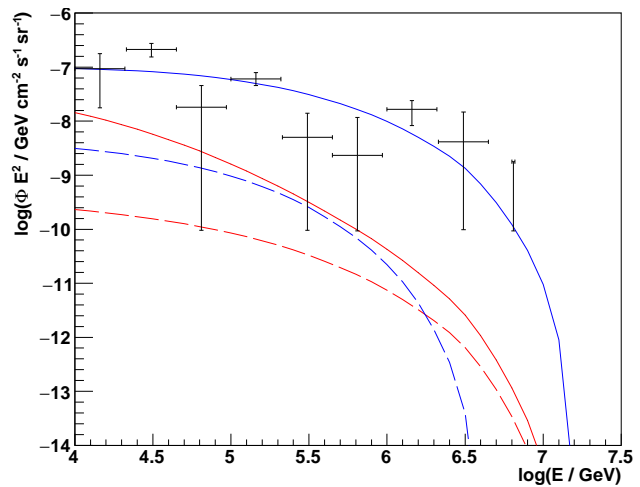}
\caption{Extragalactic diffuse neutrino background (ENB) calculated for the uniform medium surrounding the GRB (thick curves) and the non-uniform medium (thin curves) as discussed in Sect. 4.. The model without adiabatic energy losses of hadrons is shown by the solid curve and with adiabatic energy losses is shown by the dashed curves. The spectrum of the extragalactic neutrino background, measured by the IceCube Collaboration (Aartsen~et~al.~2015), is shown as error bars. }
\label{fig3}
\end{figure}

\section{Conclusion}

We consider the model for the neutrino production in the dense regions surrounding the GRBs. In this scenario, protons, accelerated in the GRB jet, escape from the acceleration site in the GRB jet to the dense medium, producing neutrinos in collisions with the background matter. It takes quite some time for relativistic hadrons to reach a dense cloud. protons are confined within the cloud for a relatively long period
producing neutrinos at a relatively low rate but for a long time.
Therefore, our model does not predict neutrino emission accompanying the specific GRBs. On the other hand, neutrinos from the surrounding cloud, are produced with large delay, forming an extended afterglow emission which cannot be directly identified with a specific GRB. Neutrino emission, produced in terms of this scenario, is expected to last for thousands of years after the initial GRB. Therefore, we conclude that the observed extragalactic neutrino background can originate in the surrounding of the GRBs which exploded a long time ago. They cannot be related to the presently observed population of GRBs at other energies.

\section*{Acknowledgments}
This work is supported by the grant through the Polish Narodowe Centrum Nauki \\
No. 2019/33/B/ST9/01904.

%\begin{thebibliography}{}

\vskip 1truecm

\leftline{Bibliography}

Aartsen, M. G. et al. 2015, ApJ 809 98

Aartsen, M. G. et al. 2017, ApJ 843 112

Abbasi, R. et al., 2021, PRD 104, 022002

Abdalla, H. et al. 2019 Nature 575, 464

Abdo, A.A. et al. 2009 Science 323, 1688

Acciari, V.A. et al. 2019a Nature 575, 455

Acciari, V.A. et al. 2019b Nature 575, 459

Ackermann M. et al., 2013, ApJS 209, 11

Ackermann M. et al., 2014, Science 343 42

Dai, Z., G., Lu, T., 1998, MNRAS, 298, 87

Heck, D., Knapp, J. et al., Technical Report 6019 (1998) Forchungszentrum, Karlsruhe

Holtzman, J.A. et al. 1992, AJ 103 691

Hurley, K. et al 1994 Nature 434, 1098

Lloyd-Ronning N. M. et al., 2019, MNRAS 488 4 5823 

Miller, B.W. et al. 1987, AJ 114, 2381

Murase, K. 2017, PRD 76 123001

Paczy\'nski, B. 1998 ApJ 494, L45

Panaitescu, A., Kumar, P., 2001, ApJ 560 L49

Piran, T. 2005 in \emph{Magnetic fields in the Universe: From Stars to Primordial Structures}, eds. E.M. de Gouveira Dal Pino et al. AIP 0-7354-0273-6/05

Razzaque, S., 2013, PRD 88 103003

Sari, R., 1997, ApJL 489 L37

Smith, N., Gehrz, R. D., Krautter, J. 1998 AJ 116, 1332

Smith, N., Gehrz, R. D., Hinz, P. M., et al. 2003 AJ 125, 1458

Thomas, J.K. et al., 2017, PRD 69, 103004 

Weaver, R. et al.,  1977, ApJ 218 377

Whitemore, B.C., Schweizer, F., 1995, AJ 109 960

Vanzella, E. et al., 2017, MNRAS 467, 389

Yost, S.A. et al., 2003, ApJ 597 459 

Waxman, E., Bahcall, J.N. 2000 ApJ 541, 707

Woosley, S.E. 1993 ApJ 405, 273

Zepf, S.E. et al., 1999, AJ 118 752

Zhang, B. 2019 \emph{The Physics of Gamma-Ray Bursts}, Cambridge University Press,
DOI: 10.1017/9781139226530

%\end{thebibliography}

\end{document}